# Ferromagnetic order beyond the superconducting dome in a cuprate superconductor

Tarapada Sarkar[1], D. S. Wei[2,3], J. Zhang[6], N. R. Poniatowski[1], P. R. Mandal[1], A. Kapitulnik[2−5], and Richard L. Greene[1,*]

[1]Maryland Quantum Materials Center and Department of Physics, University of Maryland, College Park, Maryland 20742, USA.

[2] Geballe Laboratory for Advanced Materials, Stanford University, Stanford, CA 94305, USA.

[3]Department of Applied Physics, Stanford University, Stanford, CA 94305, USA.

[4]Department of Physics, Stanford University, Stanford, CA 94305, USA.

[5]Stanford Institute for Materials and Energy Sciences (SIMES),SLAC National Accelerator Laboratory, 2575 Sand Hill Road, Menlo Park, CA 94025, USA.

[6]State Key Laboratory of Surface Physics, Department of Physics, Fudan University, Shanghai 200433, People's Republic of China.

*To whom correspondence should be addressed; email: rickg@umd.edu

**The cuprate high-temperature superconductors (HTSC) have been the subject of intense study for more than 30 years with no consensus yet on the underlying mechanism of the superconductivity. Conventional wisdom dictates that the mysterious and extraordinary properties of the cuprates arise from doping a strongly correlated antiferromagnetic (AFM) insulator (1, 2). The highly overdoped cuprates−those beyond the dome of superconductivity (SC)–are considered to be conventional Fermi liquid metals (3). Here, we report the emergence of itinerant ferromagnetic order (FM) below 4K for doping beyond the SC dome in electron-doped $La_{2−x}Ce_xCuO_4$ (LCCO). The existence of this FM order is evidenced by negative, anisotopic and hysteretic magnetoresistance, hysteretic magnetization and the polar Kerr effect, all of which are standard signatures of itinerant FM in metals (4, 5). This surprising new result suggests that the overdoped cuprates are also influenced by electron correlations and the physics is much richer than that of a conventional Fermi liquid metal.**

Many ordered phases besides AFM, such as charge order (*6–8*) and nematicity (*9*), have been found in under-doped cuprates and their relationship to HTSC have yet to be determined. Moreover, the much studied pseudogap (PG) in the p-type cuprates appears to be an unknown phase which ends at a critical doping and may play a role in the mechanism of the SC (*10*). The seemingly conventional overdoped region of the phase diagram (beyond the PG end point in p-type or beyond the Fermi surface reconstruction (FSR) in n-type) have been the subject of less thorough study, and their importance to the mechanism of SC largely dismissed. However, there are some studies of this region that suggest the physics is not that of a conventional Fermi liquid (FL). For example, in both p-type and n-type materials, the low temperature normal state transport properties are anomalous (*11, 12*). An extended range of "quantum critical" transport appears to exist from the FSR doping to the end of the SC dome (*13–15*).

The surprising new FM order that we report here in n-type LCCO raises further challenges to the conventional picture of overdoped cuprates. Hints of magnetism in overdoped p-type cuprates at temperature below 1 K have been reported previously (*16*) and FM fluctuations have been reported at higher temperatures (*17*). In this work, we provide comprehensive and robust evidence for static FM order in LCCO as will be detailed below. It should be noted that in 2007, Kopp et al. (*18*) hypothesized that the temperature dependence of the magnetic susceptibility ($x$) in overdoped $(Bi,Pb)_2Sr_2CuO_{6+\delta}$ is caused by the fluctuations of a FM phase that exists beyond the end of the superconducting dome and which competes with the d-wave superconductivity. So, it is likely that FM is a universal feature of overdoped cuprate physics. However, until our work, there has been no direct experimental evidence of static (or itinerant) ferromagnetic order associated with the $CuO_2$ planes in any cuprate. It is possible that this FM order may represent another intriguing similarity between the cuprates and twisted bilayer graphene, given that FM order has recently been found beyond the SC dome in that system (*33*).

To investigate this highly overdoped regime we measured electron-doped LCCO thin films, which can reliably be doped beyond the superconducting dome. In particular, we will focus on the non-SC dopings (x = 0.18, 0.19) where a FL-like quadratic temperature dependence of the resistivity is found at low temperatures (*19*). We observe the itinerant ferromagnetic order at temperatures below 4 K in these non-SC samples. In contrast, no ferromagnetic order is found for dopings inside the SC dome (*x* < 0.175), suggesting the existence of a ferromagnetic quantum critical point at the end of the SC dome at $x = 0.175$. This could explain the mysterious quantum critical behavior found previously near the end of the SC dome in LCCO (*20*). The evidence for FM order below 4K in overdoped (non-SC) LCCO is based on transport and magnetization measurements of numerous thin film samples grown on three different substrates. Strong confirmation of this interpretation is given by the results of a polar Kerr effect experiment on a non-SC overdoped LCCO film on a LSAT substrate.

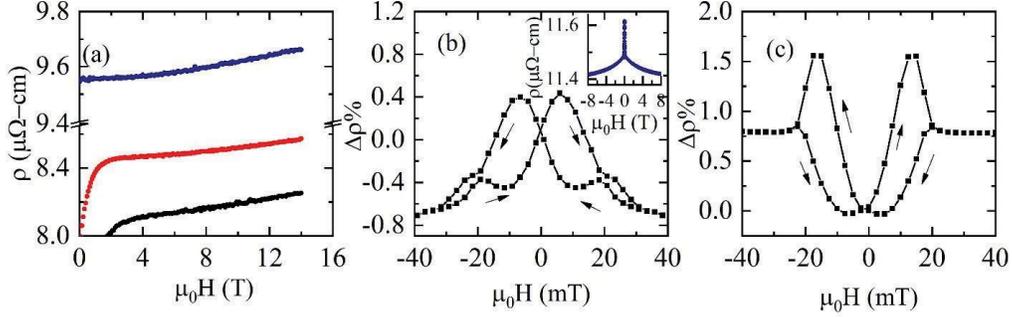

**Figure 1: Low temperature magnetoresistance across the end of the SC dome:** (a) $ab$-plane magnetoresistivity ($H \perp ab$-plane) for $x = 0.17$ ($T_c = 4$K) at 2 K, 5 K, and 10 K; (b) and (c) $ab$-plane $\Delta\rho(\%) = (\rho(H) - \rho(0))/\rho(0) \times 100$ in low field sweep from +400 Oe to -400 Oe for $H \perp ab$-plane and for $H \parallel ab$-plane respectively at 2 K for $x = 0.18$ ($T_c = 0$ K). Arrows indicate the sweeping direction of the $H$ field. Inset: higher field $ab$-plane magnetoresistance for $H \perp ab$-plane at 2 K.

In Figs 1 and 2 we present the negative transverse magnetoresistance (MR), anisotropic MR, and magnetic field hysteresis in magnetization, MR, and magneto-thermopower measurements. Figure 1 shows the low temperature transverse ($H \perp ab$-plane) magnetoresistance for both a SC (x = 0.17) and non-SC (x = 0.18) sample. The MR for x = 0.17 is positive and crosses over from linear to quadratic in field with increasing temperature, while the transverse MR for x =0.18 is negative with a strong low field hysteretic dependence below ~ 4 K. Both of these features: the negative MR and low field hysteresis below 4 K, are hallmarks of itinerant ferromagnetism (*4, 21*). Similar MR data suggestive of FM order is shown in Figs. 2a and 2b for x = 0.19, again below 4 K. As shown in Fig. 2d, we also observed hysteresis in the magneto-thermopower (MTEP) (*22*), re-affirming the presence of FM and ruling out any current heating effect as the cause of the MR hysteresis. In Fig. 2c we show a SQUID magnetization (M) study of a x = 0.19 sample which demonstrates hysteresis in the magnetization below 4 K, with a coercive field comparable to that of the MR shown in Figs. 2a and 2b. The hysteresis vanishes at 4 K as shown in Fig. S1. At 2 K the magnitude of the magnetization is approximately 0.06-0.08 $\mu_B/fu$ (formula unit). The fact that the in-plane magnetization saturates whereas the out-of-plane magnetization does not saturate (as shown in Fig. 2c) shows that the FM moments are in the plane, as confirmed by the polar Kerr effect results (Fig. 3 and SI). Furthermore, we found an anisotropic magnetoresistance (AMR), which is a well-known effect intrinsic to ferromagnets, arising due to spin orbit coupling (*4, 23, 24*). This data is shown in Figs. 1b, 1c, and S4, where the MR depends on the relative orientation of the current and the magnetization.

In Fig. 3 we show the polar Kerr effect measurements of an x =0.19 LCCO film grown on an LSAT substrate. For these high-resolution measurements, we use a zero-area loop Sagnac interferometer that nulls out any reciprocal effects such as linear birefringence and is capable of detecting tiny magneto-optic effects on the order of tens of nanoradians, as described in ref. 26. This technique has been previously used to detect time reversal symmetry breaking in triplet superconductors (*26, 27*) as well as the ferromagnetic phase transition of a thin film of SrRuO$_3$ (*5*). For all the Kerr measurements we operate at a wavelength of 1550 nm with a spot size D $\sim$ 10.6 $\mu$m and measure with an incident optical power P$_{inc}$ $\sim$ 30 $\mu$W. The details of these measurements are given in the Fig. 3 caption and the SI. These results fully confirm the transport and magnetization results and indicate the onset of long-range ferromagnetic order at $\sim$ 4K. Moreover, the magnitude of the Kerr angle $\theta_K$ is consistent with the magnetic moment found from the magnetization results (see SI).

Altogether, the negative transverse MR, anisotropic MR, and hysteretic features in MR, MTEP, and magnetization measurements and the polar Kerr effect provide compelling, unequivocal evidence for static ferromagnetic order below 4 K in overdoped LCCO. Great care was taken to ensure the observed magnetism is intrinsic to the sample (see SI section d for details), including reproducing our results for over 20 films grown on three different substrates. For example, magnetoresistance hysteresis similar to Figs. 1 and 2 for LCCO on an STO substrate is found for LCCO films on LSAT and LSGO substrates as shown in the SI, Fig. S3. Magnetization hysteresis of LCCO on LSAT is shown in Fig. S1c and the Kerr effect was done for LCCO films on an LSAT substrate.

The FM we observe in overdoped non-SC LCCO resembles that found in weak itinerant ferromagnets such as UGe$_2$ (*28*) and Y$_4$Co$_3$ (*29*), in that they also exhibit a $T^2$ temperature dependence of the resistivity. The FM order may exist above the doping $x$ = 0.19, but we were unable to prepare such films. The onset of SC at 5 K for $x$ = 0.17 prohibits a lower temperature MR study for this and lower dopings. However, we measured an $x \sim$ 0.175 film which might be SC below 1.8 K. This film shows a positive normal state MR at 1.8 K and no low field hysteretic MR (see SI Fig.S12), which means it is not FM at 1.8 K. Based on this result, the MR data in Fig.1 and a prior $\mu$SR study of LCCO (*30*), it is reasonable to predict the absence of any FM order below $x = 0.175$. We attribute the observed ferromagnetism to the hypothesized (*18*) low temperature ferromagnetic order in the copper oxide planes of overdoped cuprates.

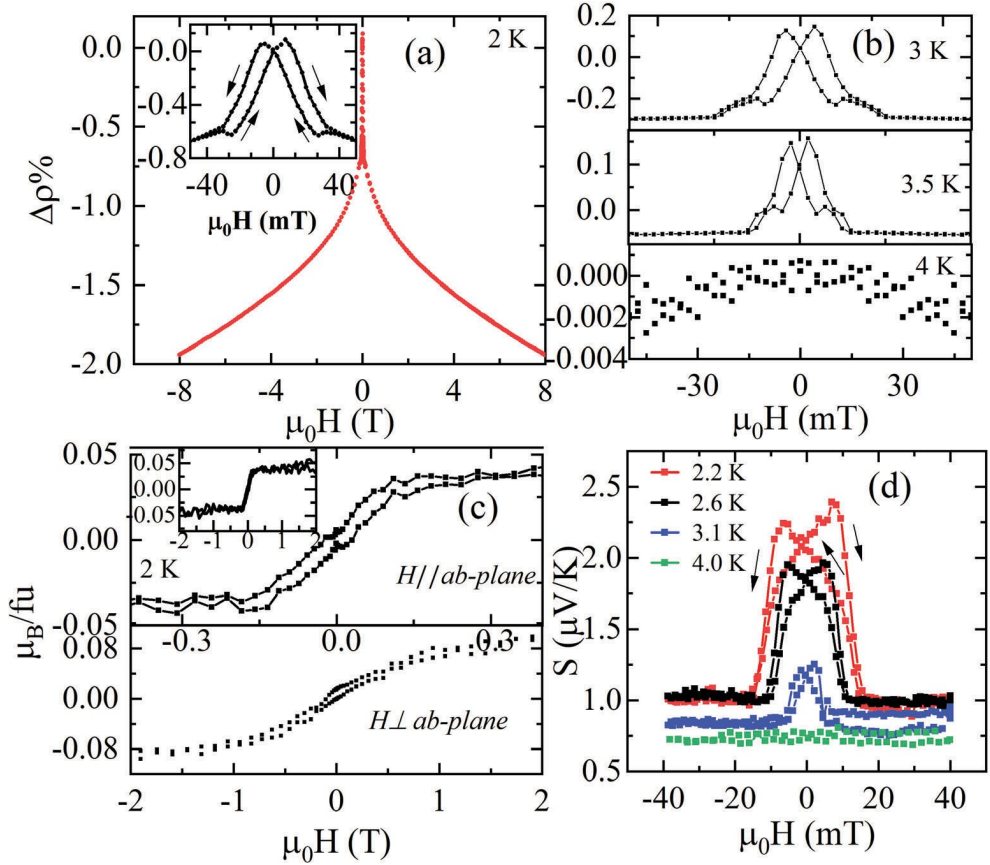

**Figure 2: Magneto-transport and magnetization for** $x = 0.19$: (a) $ab$-plane $\Delta\rho(\%) = (\rho(H) - \rho(0))/\rho(0) \times 100$ ($H \perp ab$-plane) at 2 K; inset: $\Delta\rho(\%)$ in extended view. Black arrows indicate the sweeping direction of the field; (b) low field $ab$-plane $\Delta\rho(\%)$ ($H \perp ab$-plane) at temperatures 3 K, 3.5 K, and 4 K with same sweeping direction as shown in Fig. 2(a) (the $y$-axis scale is also the same); (c) magnetization versus magnetic field with $H \parallel ab$-plane and $H \perp ab$-plane at 2 K. The substrate background is removed in these plots. Inset: low field magnetization in extended view for $H \parallel ab$-plane (see SI for more details). (d) $ab$-plane thermoelectric power with transverse sweeping field +400 Oe to -400 Oe.

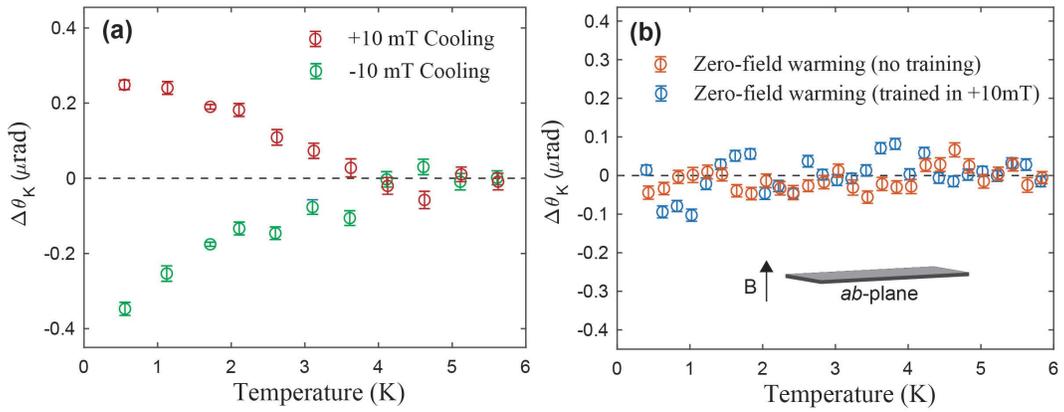

**Figure 3: Measurement of polar Kerr effect in LCCO:.** (a) Kerr angle measured in zero magnetic field after cooling down from 6K in +10 mT (red) and -10 mT (green), plotted as a function of temperature for an LCCO sample with x = 0.19 ($H_\perp ab$-plane, LSAT substrate). Data is averaged over 500 mK windows and we subtract a temperature-independent background offset of roughly 0.7 $\mu$m due to electrical and optical contributions from the instruments. Error bars in the Kerr data indicate standard deviation of the mean. An onset of Kerr signal at 4K and a complete reversal with opposite magnetic field indicates field-canted moments along the polar direction. (b) Kerr angle measured in zero magnetic field while warming after cooling down in zero magnetic field (orange) and after cooling down in +10mT (blue). Data is averaged over 200 mK windows and we subtract a temperature-independent background offset of roughly 0.7 $\mu$m. There is no discernible Kerr signal, indicating that the magnetic moments are in the ab-plane. Inset figure shows the direction of the applied magnetic field for all Kerr measurements.

A previous transport study observed quantum critical behavior of unknown origin at the end of the SC dome in LCCO based upon the scaling of the resistivity with temperature and magnetic field. The work reported the low temperature normal state resistivity to vary as $T^{1.6}$ for a doping at the end of the dome (*20*). This power law of resistivity is very close to the power law expected to arise from quantum critical FM fluctuations (*31*). In conjunction with the discovery of FM order above the SC dome described in this work, these results are suggestive of a SC/FM quantum critical point located at the end of the SC dome in LCCO.

In conclusion, our present study firmly establishes the existence of itinerant ferromagnetic order in the overdoped, non-SC, cuprate LCCO at temperatures below 4 K. This suggests the presence of a ferromagnetic quantum critical point at the end of the superconducting dome, and a resultant competition between d-wave superconductivity and ferromagnetism. This competition may play a role in other unexplained aspects of the overdoped cuprates, such as the decrease of $T_C$ beyond optimal doping and the anomalous loss of superfluid density (*32*). However, future work will be needed to elucidate more about the nature of the FM phase and its impact on the properties of overdoped cuprates. Nevertheless, this striking observation of itinerant ferromagnetism should offer new insights to address the long-standing mystery of the cuprates, and reimagine the unexplored frontiers of their phase diagram.

**ACKNOWLEDGEMENTS:** The work at the University of Maryland is supported by the NSF under Grant No. DMR-1708334, AFSOR under grant no. FA9550-14-10332 and the Maryland Center for Nanophysics and Advanced Materials (CNAM). Work at Stanford University was supported by the Department of Energy, Office of Basic Energy Sciences, under contract no. DE-AC02- 76SF00515. D.S.W. acknowledges support from the Karel Urbanek Postdoctoral Fellowship in Applied Physics at Stanford University. J.Z. acknowledges support from the Chinese Government Scholarship of China Scholarship Council. The Kerr effect experiments were funded in part by a QuantEmX grant from ICAM and the Gordon and Betty Moore Foundation through Grant GBMF5305 to Nicholas R. Poniatowski. We thank Drs. Michael Coey, Jochen Mannhart, Johnpierre Paglione, Nicholas Butch, Wesley Fuhrman, and Joshua Higgins for helpful discussions and comments on the manuscript.

**Author contributions:** R.L.G. directed the overall project. A. K. directed the research at Stanford. T.S. performed the transport and magnetization measurements and analysis. P.R.M. performed the thermoelectric measurement. T.S. and N. R. P. prepared the samples. D.S.W and J.Z. performed the Kerr effect measurements and analysis with assistance from N.R.P. R.L.G., T.S., N.R.P. and A.K. wrote the manuscript and discussed with all other authors.

**Competing financial interests** The authors declare no competing financial interests.

**SUPPLEMENTARY MATERIALS:**

www.sciencemag.org/content

Materials and Methods

Supplementary Text

Figs. S1 to S12

References

Supplementary Information

## Materials and Methods

High quality La$_{2-x}$Ce$_x$CuO$_4$ (LCCO) thin films $\sim$ 150 - 200 nm thick were grown using the pulsed laser deposition (PLD) technique on SrTiO$_3$ [100] (STO), (LaAlO$_3$)$_{0.3}$(Sr$_2$TaAlO$_6$)$_{0.7}$ [100] (LSAT), LaSrGaO$_4$ [100] (LSGO) substrates (5 $\times$ 5 mm$^2$) at a temperature of 720-750 °C utilizing a KrF excimer laser at oxygen partial pressure 230 mTorr. The films were post annealed at temperature 600-640 °C for 30-40 minutes at pressure 2$\times$10$^{-5}$ Torr. Measurements were performed on La$_{2-x}$Ce$_x$CuO$_4$ (LCCO) films for $x$ = 0.17, 0.18, 0.19 compositions (see Fig.S10 for the phase diagram). More than 20 different films have been measured by the various techniques discussed in this paper. All showed the signatures of ferromagnetic order. The LCCO targets were prepared by the solid-state reaction method using 99.999% pure La$_2$O$_5$, CeO$_5$, and CuO powders. The Bruker X-ray diffraction (XRD) of the films shows the $c$-axis oriented epitaxial LCCO tetragonal phase. The thickness of the films has been determined by using cross sectional scanning electron microscopy (SEM). The magnetotransport measurements of the films have been carried out from 2 K to 100 K in DC magnetic fields up to $\pm$ 9 T in a Quantum Design Physical

Property Measurement System (PPMS). The Hall component in the magnetoresistance is removed by adding positive sweep and negative sweep and dividing by 2. The magnetoresistance measurement is performed on patterned (50 micron channel) and unpatterned samples of sizes $1 \times 5$ mm$^2$ and $5 \times 10$ mm$^2$. The magnetization measurement was conducted in a Magnetic Property Measurement System (MPMS) equipped with a 7 T magnet. The thermopower measurement utilized a single heater technique (see Ref-4 in the main text).

## Supplementary text
## (a) Magnetization

Figure S1a shows the magnetization ($M$) vs field ($H$) data for doping $x$ =0.19 LCCO grown on STO substrate from 2 K to 10 K for $H \parallel ab$-plane. The magnetization is hysteretic below 4 K, with the feature vanishing above 4 K. This is consistent with our magnetoresistance measurements in which the hysteresis loop is seen only below 4 K. Figure S1b shows a similar hysteresis in the magnetization for $H \perp ab$-plane with a slightly higher magnetic moment than found in the in-plane measurement. Fig. S1c shows the magnetization ($M$) vs field ($H$) data for doping $x$ =0.19 LCCO grown on LSAT substrate at 2 K for $H \parallel ab$-plane. The magnetization is independent of substrates, which rules out any extrinsic magnetism coming from the STO substrate or the interface. Fig.S2 shows the in-plane magnetization of the same STO substrate as shown in Fig.S1a after removing the LCCO film by chemical etching (the sample was dipped in HNO$_3$ for 45 seconds). As shown in the inset, there is no hysteric feature, again showing that the substrate is not the origin of the observed magnetism. The $M$ vs $H$ curve at 2 K with the STO substrate background subtracted out is plotted in the main text Fig. 2c. The background signal is different for in-plane and out-of-plane measurements. A quartz holder was used for both the in-plane and out-of-plane measurements and GE Varnish was used as glue to mount the sample for the in-plane measurement. A small piece of Teflon glued with GE Varnish was also used for the out-of-plane measurement, and is why the diamagnetic background is different for in-plane and out-of-plane measurements. This difference is seen even at room temperature. These measurements show that the observed ferromagnetic order below 4 K is intrinsic to LCCO, and not a contribution from the substrate.

## (b) Magnetoresistance of LCCO grown on three different substrates

Figure S3 (a), (b) show the low field magnetoresistance LCCO with doping x=0.19 grown on LSGO and LSAT substrate at 2 K for $H \perp ab$-plane. The magnetoresistance grown on LSGO or LSAT substrate shows similar magnetoresistance as seen in LCCO grown on STO substrate for the same doping. The figure S3C shows magnetoresistance of LCCO grown on STO substrate for doping x=0.18 at temperatures 2K, 3K, 3.5K and 3.7 K.

## (c) Magneto Optical Kerr Effect

We report polar Kerr effect measurements of an electron-doped LCCO film (x=0.19) grown on an LSAT substrate. For these high-resolution measurements, we use a zero-area loop Sagnac interferometer that nulls out any reciprocal effects such as linear birefringence and is capable of detecting tiny magneto-optic effects on the order of tens of nanoradians, as described in ref. 5 in the main text. This technique has been previously used to detect time reversal symmetry breaking in the superconducting state of $Sr_2RuO_4$ (see main text ref- 27) and in heavy fermion superconductors $UPt_3$, $URu_2Si_2$, and $PrOs_4Sb_{12}$ (see ref. 1,2 and main text 28), as well as the ferromagnetic phase transition of a thin film of SrRuO3 with an out-of-plane magnetic moment (*3*). For all Kerr measurements presented we operate at a wavelength of 1550 nm with a spot size D ∼ $10.6\mu$ m and measure with an incident optical power $P_{inc}$∼ 30 $\mu$W. We apply +/-10 mT perpendicular to the ab-plane (along the c-axis) at 6K and then cool to a base temperature of T = 0.3K, taking data in 1 second intervals as the sample's temperature changes. This technique gives a qualitative measure of the magnetic moment in the c-axis and is similar to a measurement of the magnetic susceptibility. When field cooling in the presence of +10 mT, we see a finite Kerr angle develop below T = 4K, saturating at ∼ 0.3 $\mu$rad around a base temperature of T = 0.3K (Fig. X. a). Assuming a saturation Kerr signal of ∼ 1.5mrad (see e.g. $SrRuO_3$ in ref. (7), where we have ∼ 1.5 $\mu$), at ∼ 2 T (see Fig.2c), then at 10mT we expect to observe an out-of-plane Kerr signal of ∼ 1.5mrad x (10mT/2T)x($0.08\mu_B/fu/1.5\mu_B/fu$) 0.4$\mu$rad, very similar to our measured Kerr effect. Upon field cooling with -10 mT, we see that the Kerr signal changes sign with a similar magnitude as the +10 mT case. These results indicate the onset of long-range magnetic order at∼4K. We further determine that the origin of the Kerr signal is primarily from in-plane ferromagnetic order. Here antiferromagnetism is a less likely explanation because one expects the magnetization curve of an antiferromagnet to decline with temperature once the exchange energy J >> $k_B$T and the interaction overcomes the small applied field, causing the moments to anti-align. We see no such decrease in the Kerr signal as we cool to lower temperatures. Additionally, we cooled the sample in both zero field and +10 mT, and then brought the field to zero and measured upon warming from T = 0.3K to 6K (Fig X. b). Here we see no signal change at 4K, indicating that there is no spontaneous magnetic order out of the plane in the c-axis. This provides additional evidence that the magnetic order is in the ab-plane.

## (d) Intrinsic origin of the ferromagnetism

Ferromagnetism in a thin film ($\sim$ 200 nm) grown on a substrate can sometimes originate from extrinsic causes (*4*). There are several possible sources of extrinsic ferromagnetism, which we have systematically ruled out based on the following observations:

1. The magnetism is not an interfacial effect of LCCO with the STO substrate. We have reproduced the magnetization and transport results reported in the main text on STO on over 20 films grown on LSAT and LSGO substrates (as shown in Figs. S1 and S3), as discussed in parts (a) and (b) of this SI.

2. The hysteresis in a SQUID magnetometry measurement is not due to contamination from handling the sample with magnetic tweezers, as this would not lead to the observed hysteresis in magnetoresistance and magnetothermopower. Moreover, the hysteresis loop is independent of the sweeping rate of the field, which rules out any heating effect.

3. The magnetism is not caused by magnetic impurities (e.g. Fe, Co, Mn) in the film or the substrate. Our measurements rule this out for the following reasons:

   (a) The magnitude of the magnetic moment in our LCCO film is too large (0.08 $\mu_B$/fu) to be caused by magnetic impurities. A moment this large would require more than 2% of ferromagnetic impurities (like Fe, Co, Mn), which far exceeds the purity of our PLD materials.

   (b) We have measured the magnetic moment of the film after subtracting out the substrate contribution (Fig. 2c in the main text, Fig. S2).

   (c) The slightly lower doped $x$ = 0.17 shows no ferromagnetic-like magnetoresistance, even though these films are prepared from same source of metal oxide, and thus would ostensibly harbor the same impurities.

(d) To the authors' knowledge, all magnetic impurities related to weak ferromagnetism have a high Curie temperature (above 300 K) and the saturation magnetism does not change with temperature significantly (*4, 5*). In contrast, the magnetism in LCCO is seen at very low temperature (below 4 K).

4. LCCO did not show any structural change in the X-Ray Diffraction (XRD) pattern of the higher doped sample as shown in Fig. S6. The XRD pattern looks very similar throughout the doping range studied. A recent Cu L-edge resonant inelastic X-ray scattering study (*6*) of LCCO shows that Cu is in the 2+ state beyond the superconducting dome and Ce is known to be in 4+ states (*8*) in electron-doped cuprates. The $Ce^{4+}$ and $La^{3+}$ are unlikely to be ferromagnetic (no unpaired electrons and no evidence found to date). However, we cannot rule out a small amount (1% or less) of $Ce^{3+}$. But, we would not get our large measured moment from a few percent of $Ce^{3+}$ in isolation, or combined with oxygen.

Therefore, we attribute our observed ferromagnetism to an intrinsic property of the copper spins in a metallic system, i.e., an itinerant ferromagnet.

## (e) Angular magnetoresistance

The magnetoresistance of pure ferromagnetic materials has been intensively studied since the early 1960s. The ferromagnetic magnetoresistance consists of two parts, the anisotropic magnetoresistance (AMR) (see main text Ref-4, 24,25) and a small contribution from domain walls. The anisotropic magnetoresistance is a small effect depending on the relative orientation of current and magnetization, and is intrinsic to ferromagnets. It is consequence of spin orbit coupling and d band splitting, commonly seen in 3d metals or Permalloy (see main text Ref-25). In the Figure S4, LCCO shows positivive magnetoresistance when currenta and field are in the same direction in the plane. However in-plane magnetoresistance is negetive when the field is perpendicular to the plane that is perpendicular to currect (I) diirection. This anisotropic magnetoresitance in the LCCO is atributed to spin orbit coupling, which strongly suggests FM order in LCCO.

## (f) High temperature magnetoresistance

In Fig. S5, we show the higher temperature out-of-plane magnetoresistance of one $x =0.18$ LCCO sample. The MR of this sample is negative at temperatures below 50 K. At 50 K, the magnetoresistance is negative below 6 T and becomes positive at higher fields. Above 70 K the magneoresistance becomes fully positive, as shown in the inset of Fig. S5. The theory for magnetoresistance in ferromagnetic metals below the Curie temperature is well understood (7), and the negative magnetoresistance we observe above the Curie temperature in a ferromagnetic metal is not common in conventional systems. Experimentally, negative magnetoresistance is seen in several ferromagnetic transition metals with low carrier density (see Ref 19,20 in the main text), which has been explained by spin fluctuations above the Curie temperature. Cuprates are also a low carrier system leading us to speculate that the negative magnetoresistance is due to ferromagnetic spin fluctuations above the ferromagnetic transition temperature.

## (g) Low temperature Hall effect

In Fig. S6a, we show the Hall resistance measured on a lithographically patterned sample in a Hall bar geometry at 2 K. The Hall resistance at very low field exhibits a hysteresis loop because there is mixing of $R_{xy}$ and $R_{xx}$ components of the resistivity. Figure S5a shows that at zero field $R_{xy}$ is not zero, which indicates the presence of the $R_{xx}$ component. The Hall resistance in a ferromagnetic metal can be written as $R_{xy} = R_0 B + 4\pi R_s M$ (7), where the first term is the ordinary Hall effect, and the second term is the anamalous Hall effect due to the sample magnetization associated with spin orbit coupling (skew scattering) and side jumps (7). In LCCO we find the total Hall resistance, $R_T = R_{xx} + R_{xy}$, in which the first term is the dominant term at low field ( < 1000 Oe). This $ab$-plane magnetoresistance has masked the anomalous Hall component ($4\pi R_s M$ ). We removed the $R_{xx}$ contribution to the Hall resistance by subtracting the negative field Hall resistance from the positive field Hall resistance as shown in Fig. S6b. The corrected Hall resistance at 2 K is shown in the insert of Fig. S6b, which features the hysteresis loop in the first quadrant. This indicates the presence of a small anomalous Hall effect at 2 K, which is not found at 4 K.

## (h) Magnetoresistance in other oxides

To further confirm that the low field magnetoresistance is not an artifact , we plot the low field magnetoresistance of conducting Nb-SrTiO$_3$ and oxygen reduced SrTiO$_3$ (9) metal oxides in Fig. S9 (the Nb-SrTiO$_3$ is sourced from CrysTec, GmbH, Germany). We see no anomalous negative MR or low field hysteretic behavior again confirming that the low field magnetoresistance described in the main text is linked to the LCCO thin film rather than any measurement artifact. The substrates are heated and annealed under the same condition as during the preparation and annealing of the LCCO films.

## (i) Resistivity vs. temperature

In Fig. S8a we show that the $ab$-plane resistivity follows a $T^2$ power law from 2 K to 30 K for dopings $x = 0.18, 0.19$. In Fig. S8(b), we show the $ab$-plane resistivity vs temperature for the doping $x = 0.17$, which is linear in temperature as previously reported, with a superconducting transition temperature $\sim 4$ K (see Ref 1 in main text).

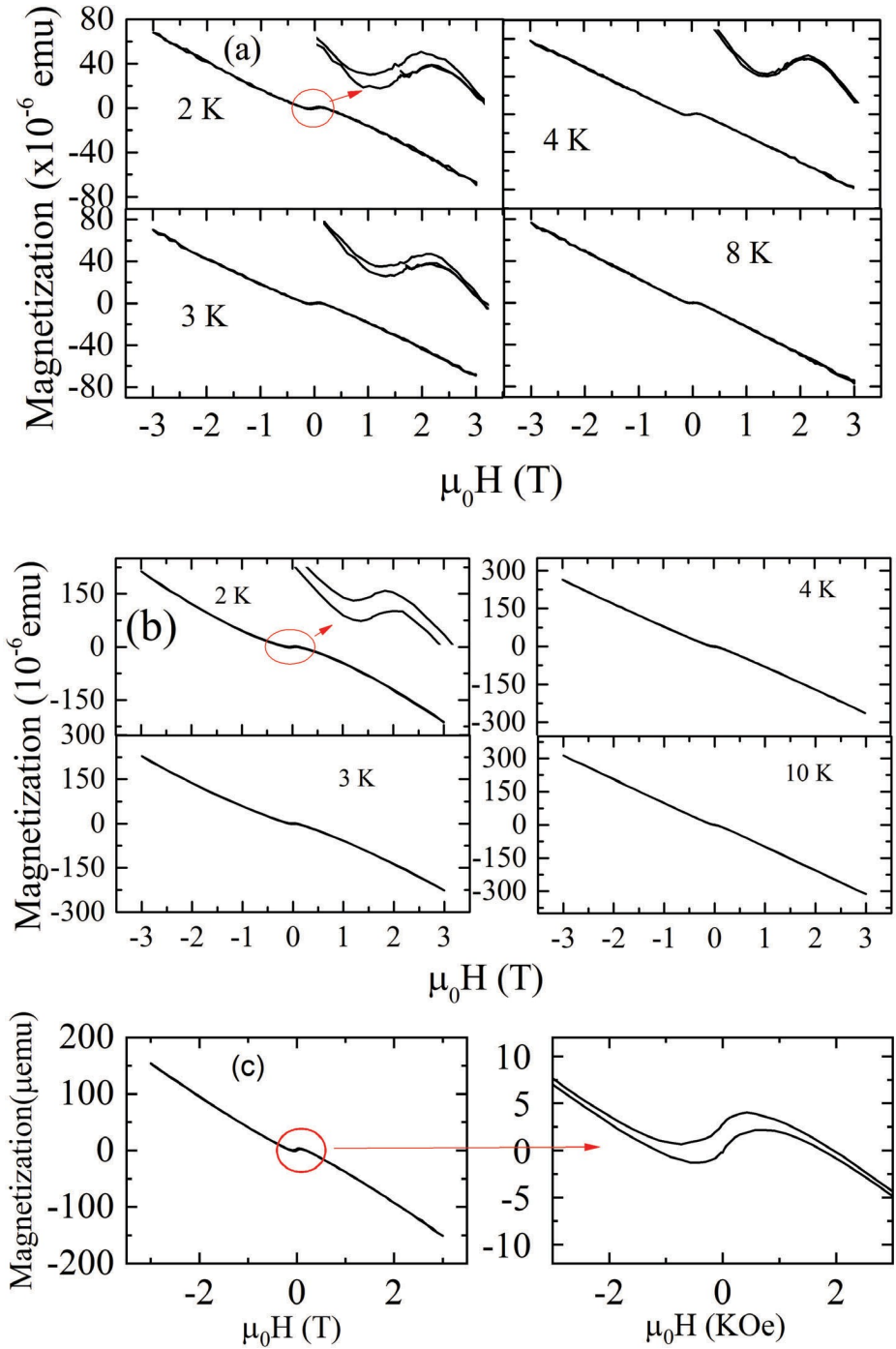

**Figure S1: Magnetization vs field:** (a) and (b) show the $M$ vs $H$ of an LCCO film grown on an STO substrate (5x2 mm$^2$ (100)) with $H \parallel ab$-plane and $H \perp ab$-plane respectively; Inset: low field zoom with $x$-axis ($+3 KOe$ to $-3KOe$) and $y$ axis ($+5$ $\mu$emu to -5 $\mu$emu; scale for all insets are same). (c) $M$ vs $H$ of an LCCO film grown on an LSAT substrate (5x2 mm$^2$ (100)) with $H \parallel ab$-plane at 2 K.

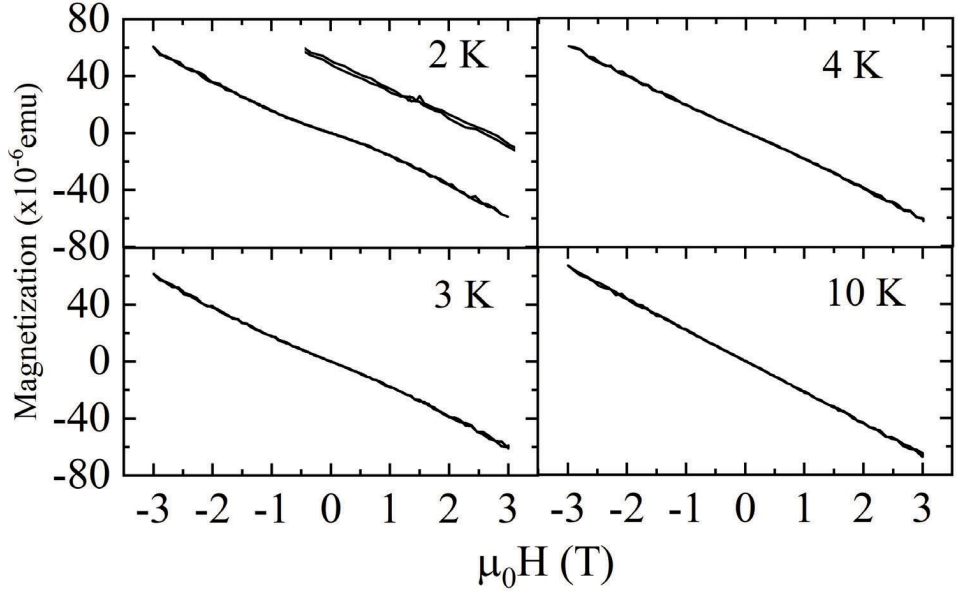

**Figure S2: Magnetization vs field:** $M$ vs $H$ with $H \parallel ab$-plane for LCCO grown on STO substrate after removing the film by chemical etching. Inset: low field zoom with $x$-axis ($+3\,KOe$ to $-3\,KOe$) and $y$ axis ($+5\,\mu$emu to $-5\,\mu$emu) as in Fig.S1.

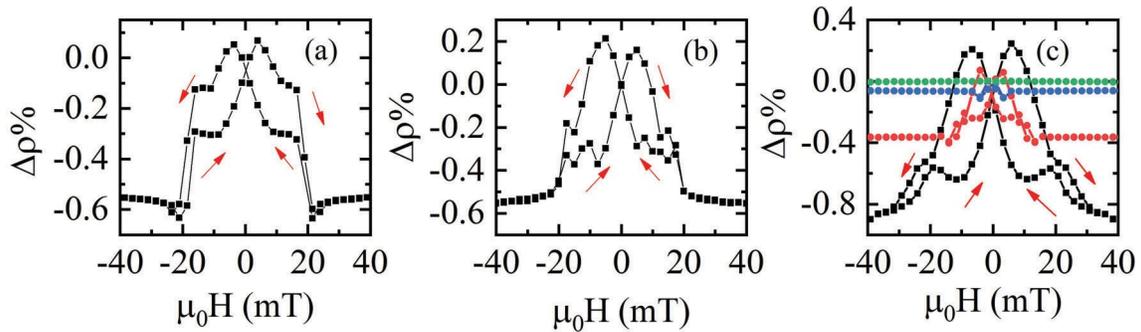

**Figure S3: Low field magnetoresistance on different substrates:** Transverse ($H_{\perp}\,ab$-plane) $\Delta\rho(\%) = (\rho(H)-\rho(0))/\rho(0) \times 100$ vs. magnetic field of LCCO for doping $x = 0.19$. grown on LSGO (a), LSAT (b) at 2 K and for doping x=0.18 on STO substarte at 2 K (black), 3 K (red), 3.5 K (blue) and 3.7 K (Green).

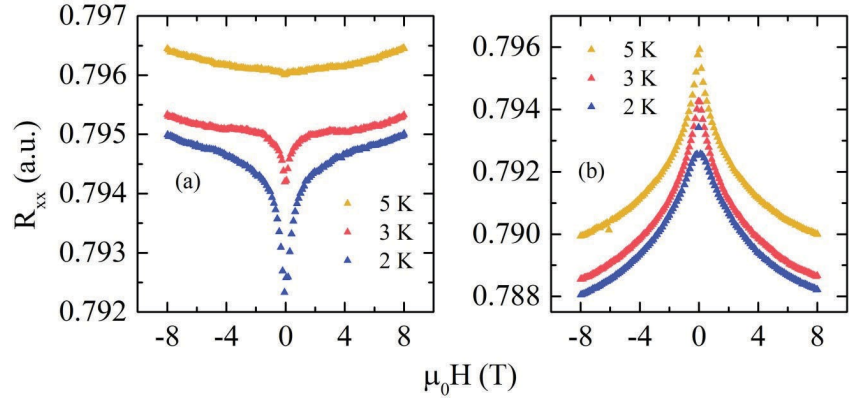

**Figure S4: Anisotropic magnetoresistance for *x* = 0.18:** (a) *ab*-plane magnetoresistance measured with field in plane and parallel to current direction (*H* ∥ *I*); (b) *ab*-plane transverse magnetoresistance (*H* ⊥ *ab*-plane) with field perpendicular to current direction (*H* ⊥ *I*). The low field data is not plotted here for clarity (see Fig. 1 and Fig. S1 for the low field magnetoresistance).

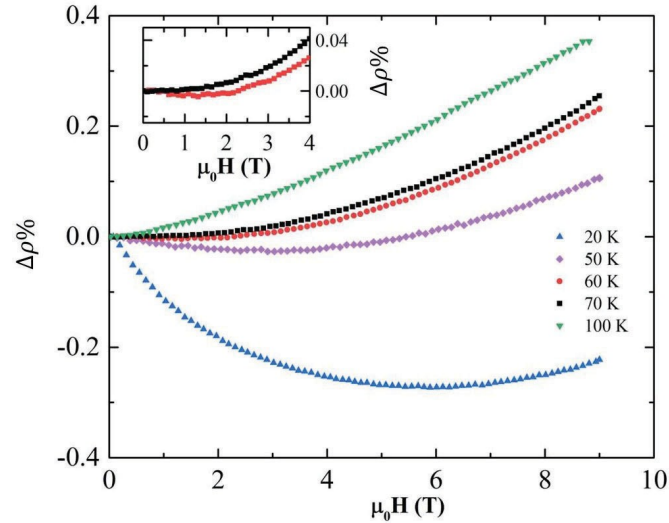

**Figure S5: High temperature magnetoresistance** Transverse $\Delta\rho(\%)$ for *x* =0.18 at various temperatures; Inset: zoom at 60 K (red) and 70 K (black).

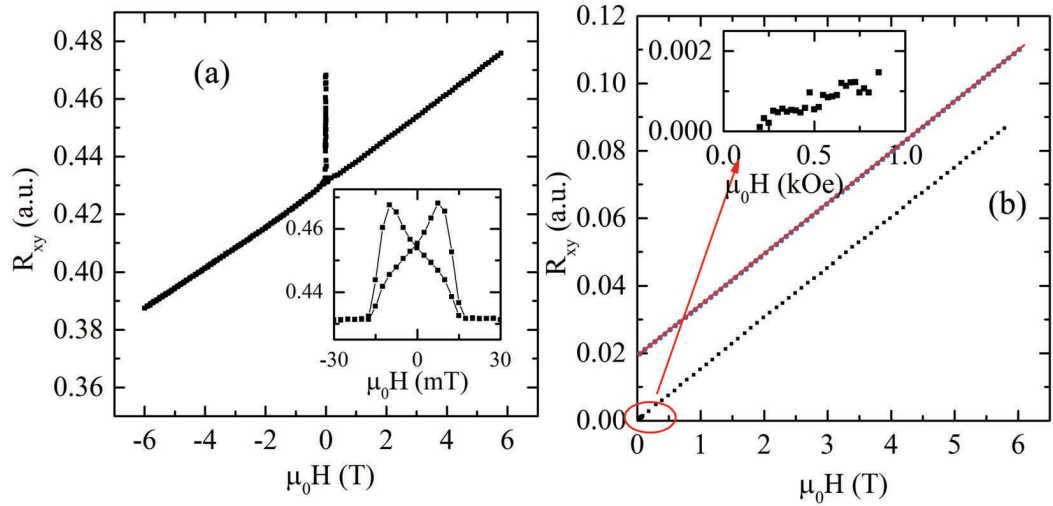

**Figure S6: Anomalous Hall Effect** (a) Hall resistance of $x$ = 0.18 at 2 K. Inset: expanded view of the low field region. (b) Hall resistance after subtracting the magnetoresistance at 2 K (black) and 4 K (blue-offset) with linear fitting (red). The inset shows the anomalous Hall component in the positive quadrant at 2K obtained by subtracting out the large $R_{xx}$ component.

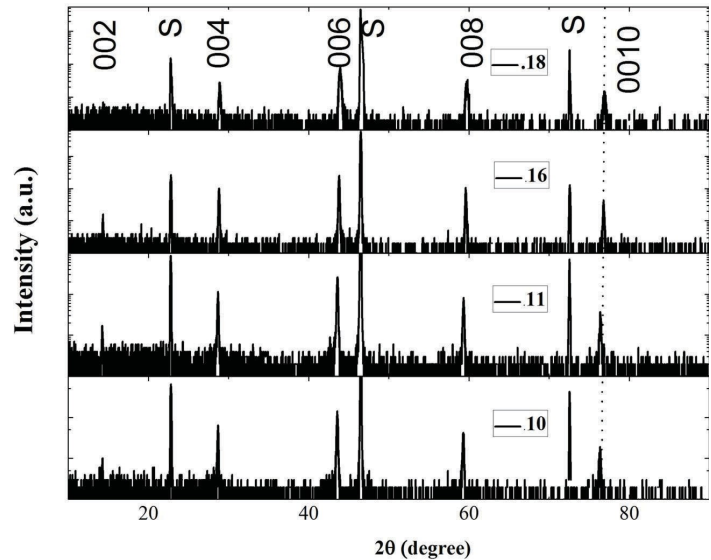

**Figure S7: X-ray Diffraction** (a) X-ray Diffraction of La$_{2-x}$Ce$_x$CuO$_4$ grown on SrTiO$_3$ (STO) substrate. The S indicates substrate peak. Dopings are indicated in each panel.

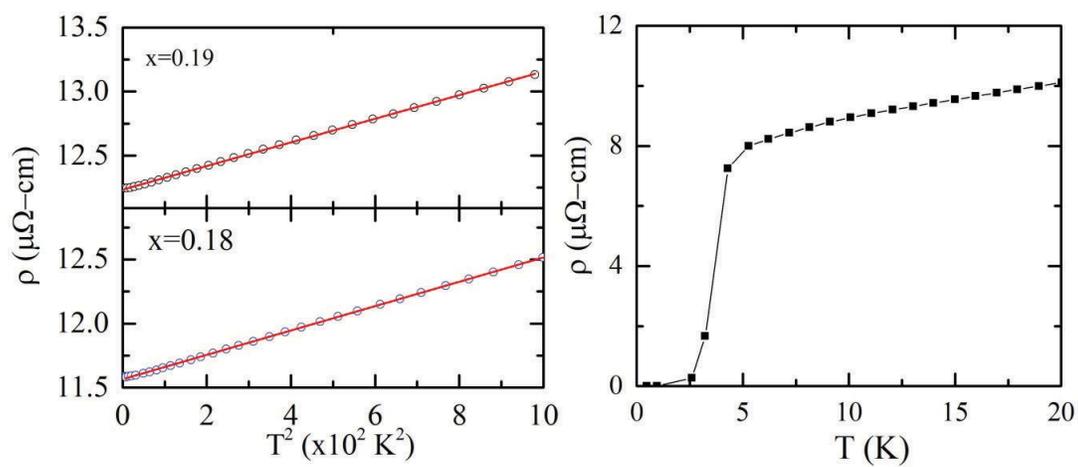

**Figure S8: Resistivity vs Temperature with doping** The resistivity ($\rho$) vs $T^2$ for dopings $x = 0.18$ and $x = 0.19$ (left panel) and the resistivity ($\rho$) vs T for doping $x = 0.17$ (right panel).

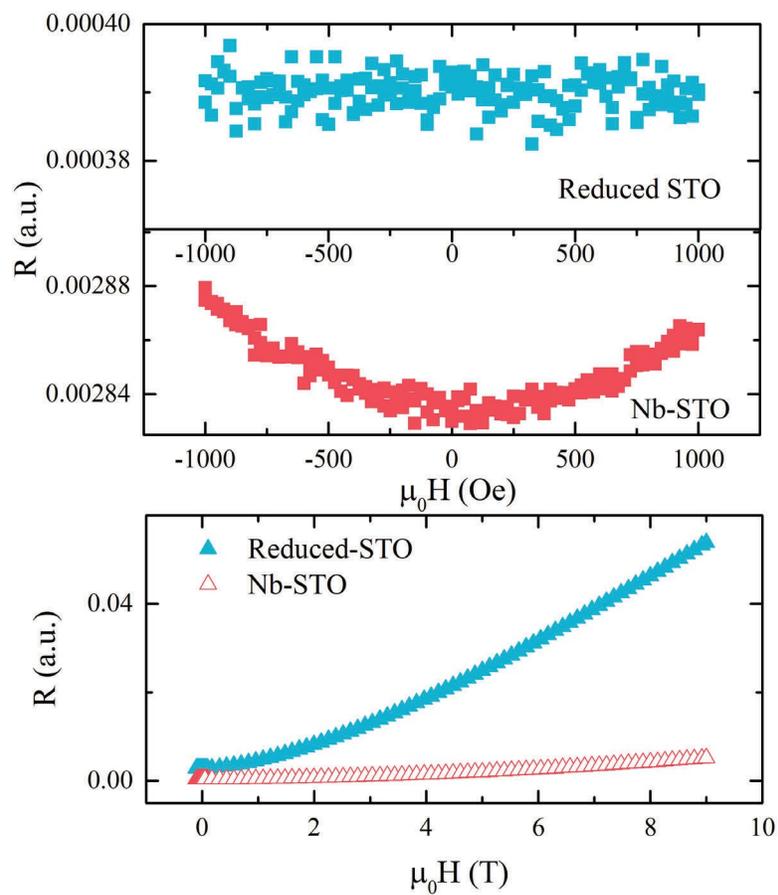

**Figure S9: Magnetoresistance calibration:** Low field $ab$-plane magnetoresistance ($H \perp ab$-plane) for reduced STO (top panel) and Nb-STO (middle panel). Lower panel shows the measurement up to 9 T.

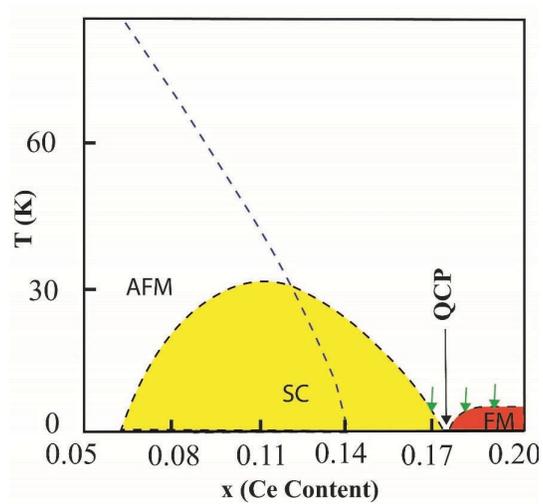

**Figure S10: Temperature vs. doping phase diagram:** Schematic phase diagram of $La_{2-x}Ce_xCuO_4$. The dotted blue line represents the region in which AF (long or short range) order is observed below doping $x = 0.14$. The yellow region represents the superconducting phase and the red region beyond the SC dome represents the itinerant FM phase found from this work. The black arrow indicates the possible FM QCP. The green arrows indicate the three dopings studied in this work.